# LANGMUIR

**Article**

# Morphology of Evaporating Sessile Microdroplets on Lyophilic Elliptical Patches

José M. Encarnación Escobar,[†,#] Diana García-González,[†,‡,#] Ivan Dević,[†] Xuehua Zhang,[*,§] and Detlef Lohse[*,†,‖]

[†]Max Planck Center Twente for Complex Fluid Dynamics, JM Burgers Center for Fluid Dynamics, Mesa+, Department of Science and Technology, University of Twente, Enschede 7522 NB, The Netherlands

[‡]Max Planck Institute for Polymer Research, Mainz 55128, Germany

[§]Department of Chemical & Materials Engineering, University of Alberta, Edmonton, Alberta AB T6G 2R3, Canada

[‖]Max Plank Institute for Dynamics and Self-Organization, Göttingen 37077, Germany

**ABSTRACT:** The evaporation of droplets occurs in a large variety of natural and technological processes such as medical diagnostics, agriculture, food industry, printing, and catalytic reactions. We study the different droplet morphologies adopted by an evaporating droplet on a surface with an elliptical patch with a different contact angle. We perform experiments to observe these morphologies and use numerical calculations to predict the effects of the patched surfaces. We observe that tuning the geometry of the patches offers control over the shape of the droplet. In the experiments, the drops of various volumes are placed on elliptical chemical patches of different aspect ratios and imaged in 3D using laser scanning confocal microscopy, extracting the droplet's shape. In the corresponding numerical simulations, we minimize the interfacial free energy of the droplet, by employing Surface Evolver. The numerical results are in good qualitative agreement with our experimental data and can be used for the design of micropatterned structures, potentially suggesting or excluding certain morphologies for particular applications. However, the experimental results show the effects of pinning and contact angle hysteresis, which are obviously absent in the numerical energy minimization. The work culminates with a morphology diagram in the aspect ratio vs relative volume parameter space, comparing the predictions with the measurements.

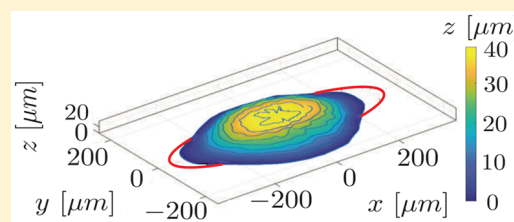

## INTRODUCTION

The use of patterned surfaces to control the behavior of liquid drops is not only a recurrent phenomenon in nature but also a useful tool in various industrial and scientific applications. For instance, the observation of nature inspired the use of patterned surfaces for water harvesting applications[1] as well as the fabrication of antifogging[2] and self-cleaning materials.[3] The geometry of droplets and the substrates in which they lie can affect their adhesion, as can their evaporation and other important properties.[4−6] The interest in wetting motivated by applications covers a wide range of scales and backgrounds from microfluidics[7−9] to catalytic reactors,[10] including advanced printing techniques,[11,12] improved heat transfer,[13,14] nanoarchitecture,[15−17] droplet-based diagnostics[18] and antiwet-antiwetting surfaces.[19−21] Aside from all the practical significance, we have to add the interest in wetting fundamentals,[22] including contact angle hysteresis and dynamics,[23−25] contact line dynamics,[26] nanobubbles and nanodroplets,[27−29] spreading dynamics,[30,31] and complex surfaces.[32,33]

Due to the complexity of the field, most previous studies have restricted themselves to considering geometries with constant curvature as *straight* stripes or *constant curvature* geometries, namely circumferences. In this work, we will study the behavior of evaporating drops on lyophobized substrates that have lyophilic elliptical patches. The elliptical shape for the patches is chosen as a transitional case between a circular patch[34] and a single stripe,[35−38] having the uniqueness of a perimeter with nonconstant curvature. When a drop is placed on a homogeneous substrate, the minimization of surface energy leads to a spherical cap shape. This is traditionally described by the Young−Laplace, Wenzel, or, for pillars or patterns with length scales much smaller than the drop, Cassie−Baxter relations,[39−42] which reasonably apply to homogeneous substrates or substrates with sufficiently small and homogeneously distributed heterogeneities,[43−45] which are not the focus of this study.

Our previous work (Dević et al.[46]) based on Surface Evolver and Monte Carlo calculations showed that when a droplet rests on an elliptical patch, four distinguishable morphologies are found, depending on the volume of the drop and the aspect ratio of the patch. These morphologies were termed[46] A, B, C and D; see Figure 1. When a large enough droplet evaporates on an elliptical patch, the first morphology found is D. In this case, the droplet completely wets the ellipse with a part of its contact line outside the ellipse and the rest pinned to the contour of the patch. After a certain volume loss, the droplet adopts either morphology B or C, depending on the geometry of the patch. In morphology B, a part of its contact line remains outside the patch while the rest is already inside the patch. In contrast, the contact line of a droplet adopting morphology C follows the perimeter of the ellipse. The final morphology to be found for evaporating drops is morphology A, characterized by a part of its contact line following the perimeter of the ellipse and the rest of the contact line lying inside it (Figure 2).

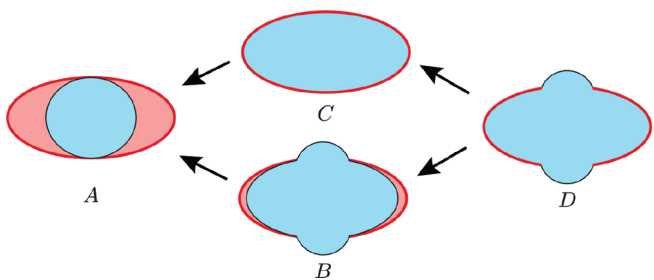

**Figure 1.** Scheme of the morphologies of a droplet as seen from the top as expected from the calculations by Dević et al.[46]

## ■ METHODS

In this paper, we perform corresponding experiments and numerical simulations, performed again with Surface Evolver using the experimental parameters. Both the experiments and the calculations are performed for Bond numbers below unity to avoid the effects of gravity. Moreover, the experiments are performed at room temperature with a relative humidity of 38 ± 2% in a closed and controlled environment, ensuring that the evaporation driven volume change is slow enough so that the droplets evolve in quasi-static equilibrium. The effects of humidity, evaporative cooling, and evaporation driven flows do not have any important effects. However, for heated or cooled substrates or more volatile liquids this situation may change. Unlike the gravitational and evaporative effects, the effects of the inhomogeneities of the substrate —like pinning and contact angle hysteresis— are unavoidable in our experiments while they are absent from the Surface Evolver calculations.

**Preparation of Substrates with Lyophilic Elliptical Patches.** The patched substrates were prepared via photolithography followed by chemical vapor deposition (CVD) of a mono layer of trimethylchlorosilane (TMCS). The 100 × 100 mm² chromium on glass photomask used was designed with an array of ellipses of aspect ratios varying from 0.3 to 1 and sizes varying from $a = 320$ μm to $b = 2500$ μm (see Figure 3a) and fabricated in the MESA+ Institute clean-room facilities of the University of Twente, using laser writing technology. The substrates were glass slides (20 × 60 mm²) of 170 μm thickness, which is optimal for confocal microscopy. The photolithography steps of the fabrication, outlined in Figure 3b, were performed in a clean-room environment. First, the substrates were precleaned in a nitric acid bath (NOH₃, purity 99%) followed by water rinsing and nitrogen drying. Subsequently, we dehydrated the substrate (120 °C, 5 min). After dehydration, we spin-coated the photoresist (Olin OiR, 17 μm), prebaked it (95 °C, 90 s), and proceeded with the alignment of the photomask and UV irradiation (4 s, 12 mW/cm²). Once the photoresist was cured, we developed and postbaked it (120 °C, 10 min). Finally, the substrates were taken outside the clean-room environment for the CVD (2 h, 0.1 MPa) of a TMCS monolayer (Sigma-Aldrich, purity ≥99%) to lyophobize the exterior of the ellipses. The photoresist was later stripped away by rinsing the substrates with acetone, cleaned with isopropanol, and dried using pressurized nitrogen.

**Experimental Data Acquisition and Analysis.** For all the measurements, we used droplets of ultrapure (Milli-Q) water dyed with Rhodamine 6G at a concentration of 0.2 μg/mL. The droplet was deposited covering the lyophilic patch and imaged by laser scanning confocal microscopy (LSCM) in three dimensions during the evaporation process. After deposition of the droplet, the substrate was covered to avoid external perturbations.

In Figure 4, we show four examples of three-dimensional reconstructions of a stack of scans with increasing heights taken under LSCM. For each scan, we detected the surface of the drop using a threshold algorithm, allowing for the three-dimensional reconstruc-tion of the drop. Using this data, we calculate the local contact angle at every point detected on the contact line. For the measurement of the contact angle $\theta$, we extract the height profile along the contact line. This can be achieved, as sketched in Figure 2, by identifying the tangent to each point of the contact line at an angle $\phi$ and finding the points belonging simultaneously to its normal plane $S$ and to the surface of the droplet.

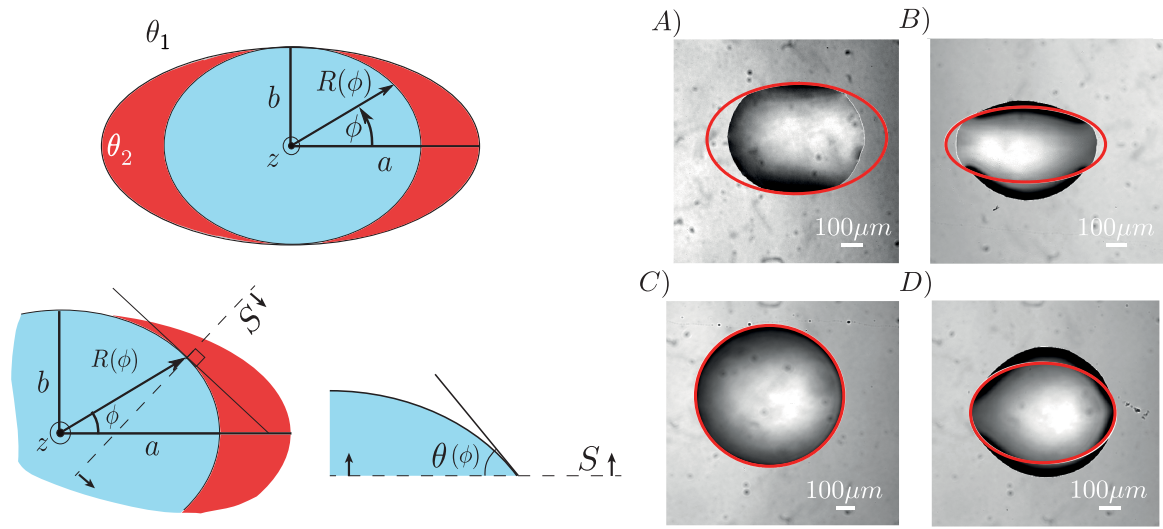

**Figure 2.** Left: Coordinate system employed in this paper. $\theta_1$ and $\theta_2$ are, respectively, the Young's angles of the lyophobic (white) and lyophilic (red) regions. $R(\phi)$ is the distance from the center of the ellipse to the contact line of the droplet (blue), and $a$ and $b$ are the major and minor axis, respectively. $S$ indicates the vertical section. Right: Experimental results. Top view images were taken during the evaporation of various droplets on ellipses with different sizes and aspect ratios; the red contours represent the elliptical patches on the surface. (A) Morphology A, droplet on an ellipse of aspect ratio $b/a = 0.61$ and semimajor axis $a = 512 \pm 16$ μm. (B) Morphology B, droplet on an ellipse of aspect ratio $b/a = 0.43$ and semimajor axis $a = 392 \pm 20$ μm. (C) Morphology C, droplet on an ellipse of aspect ratio $b/a = 0.98$ and semimajor axis $a = 410 \pm 20$ μm. (D) Morphology D, droplet on an ellipse of aspect ratio $b/a = 0.69$ and semimajor axis $a = 411 \pm 20$ μm.

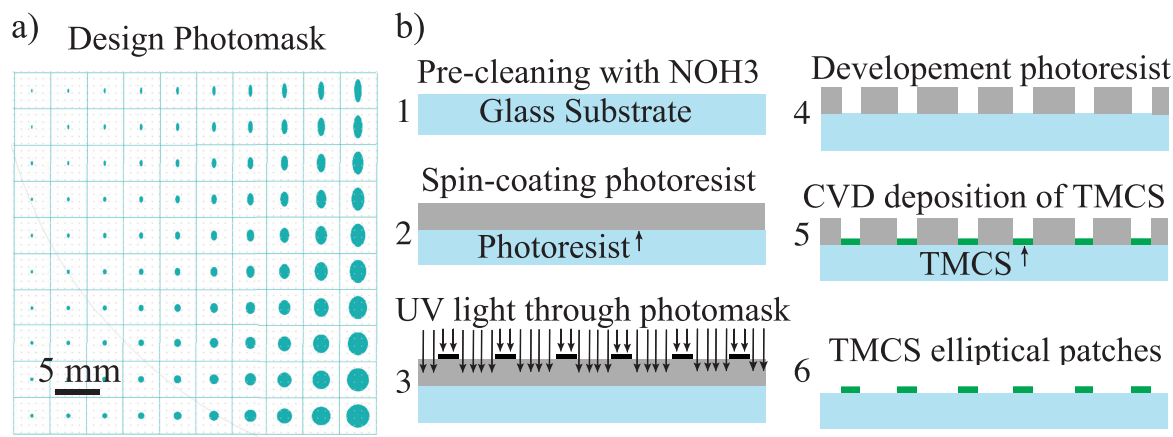

**Figure 3.** (a) Array of ellipses of aspect ratios varying from 0.3 to 1 and sizes varying from $a = 320\ \mu m$ to $b = 2500\ \mu m$ fabricated on the photomask. (b) Simplified steps of the substrates' fabrication in the order indicated by the numbers.

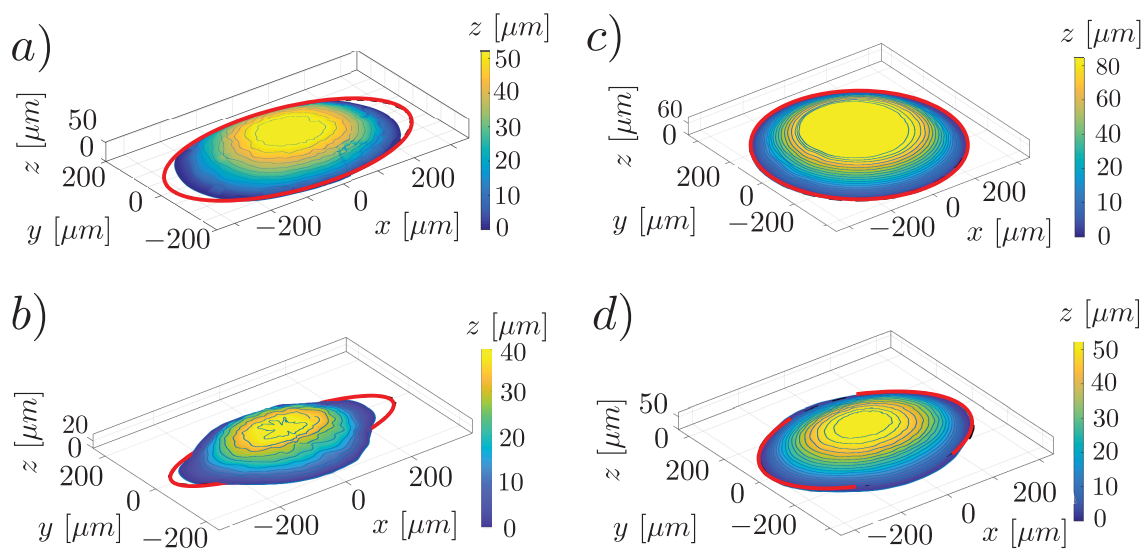

**Figure 4.** Three-dimensional reconstruction of the shape extracted from the LSCM data collected for four different droplets adopting morphologies A, B, C, and D, respectively, as performed for the extraction of the contact angle along the contact line with the shape of the elliptical patch, highlighted in red lines. Repeatability is subjected to the initial position of the droplet and pinning of the contact line which can affect the symmetry of the shape as well as delay the transition between phases as compared to the predictions. In Figure 6, all the morphologies observed during the experiments are shown.

We extracted three-dimensional images of evaporating water droplets using LSCM. From the three-dimensional data we measured the local contact angle along the three-phase contact line through the azimuthal angle $\phi$. Figure 2, right, shows top views of various evaporating droplets as examples of each of the morphologies introduced previously. The lyophilic elliptical islands of the substrate are highlighted by red curves.

We adopted a cylindrical coordinate system with its origin at the center of the ellipse. The polar axis was fixed to be in the direction of one of the major semiaxes $a$ as shown in Figure 2 (left). We set the large semiaxis ($a$) of the ellipse as the characteristic length scale for this system and the aspect ratio of the ellipses $b/a$ to characterize the geometry of the patch. The relative volume of the droplets is normalized as $V/a^3$.

**Calculations.** We compute the surface energy minimization using Surface Evolver, a free software package used for minimization of the interfacial free energy developed by Brakke,[47] to extract the droplet's shape and local contact angle as in the previous work by Dević et al.[46] The initial shape of the droplet and the characteristic interfacial tensions of the surfaces are given to Surface Evolver as an input. The software minimizes the surface energy by an energy gradient descent method. Since hysteresis is not captured by our simulations, we compared each group of experimental results with two different calculations: one considering the receding contact angles for the two regions and the other considering the advancing contact angle for the lyophilic part.

To perform the calculations, we measured the contact angles of both the lyophilic and lyophobic parts of our patches. To do this, we treated two separated substrates homogeneously in the same manner as in these two regions. We measured the advancing and receding contact angles on both substrates. For the lyophobic (subscript 1) and lyophilic surfaces (subscript 2), the advancing contact angles measured were $\theta_{a1} = 85 \pm 3°$ and $\theta_{a2} = 33 \pm 4°$, respectively, and the receding contact angles were $\theta_{r1} = 49 \pm 3°$ and $\theta_{r2} = 15 \pm 4°$, respectively. During the experiments, we observed variations between 5% and 10% of the contact angle due to occasional pinning events.

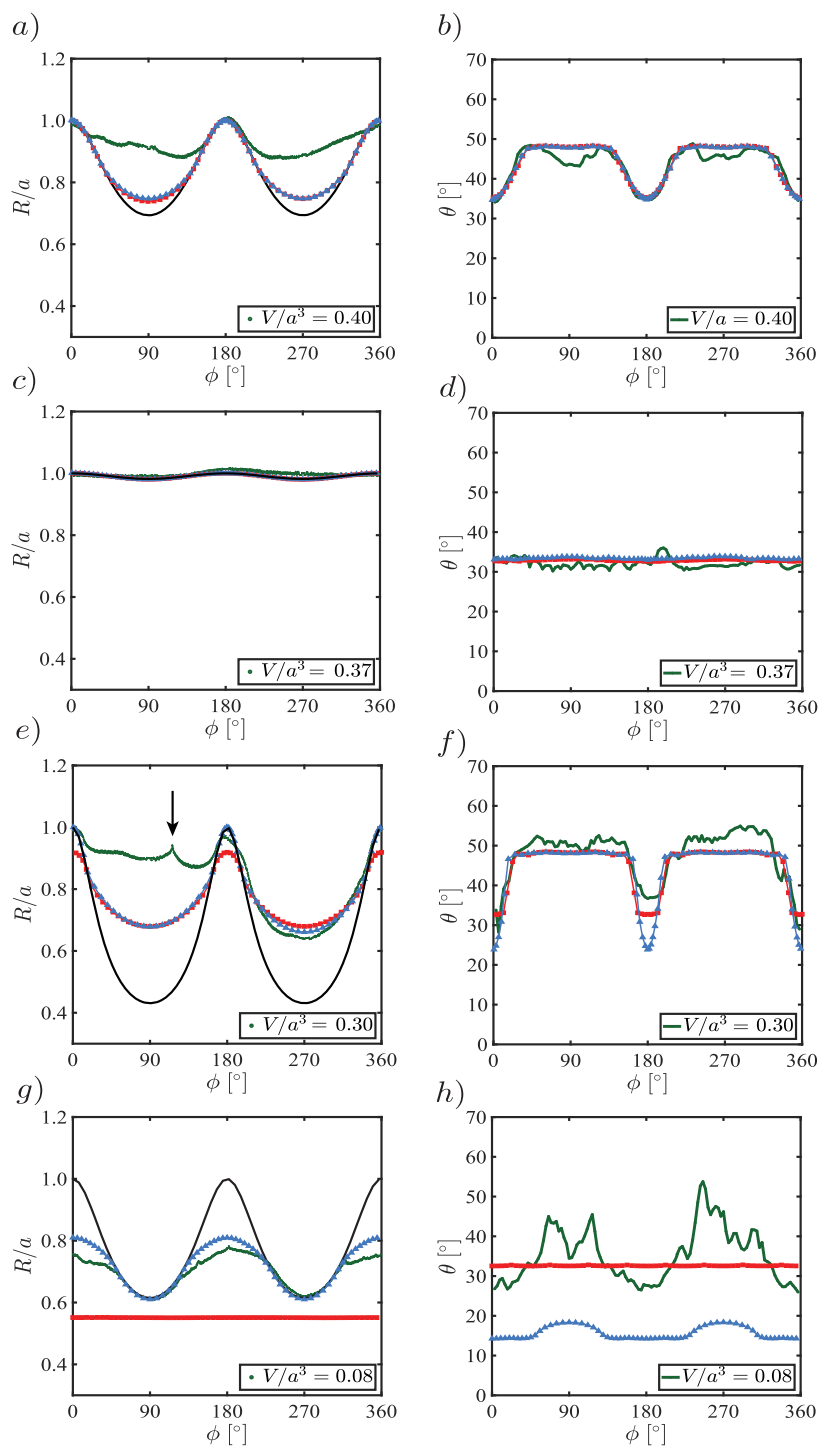

**Figure 5.** Normalized footprint radius $R/a$ and contact angle $\theta$ along the azimuthal coordinate $\phi$ (as defined in Figure 2). Experimental results for the four different morphologies (green). From top to bottom: morphology D ($V/a^3 = 0.40$, $b/a = 0.69$, and $a = 411 \pm 20$ $\mu$m); morphology C ($V/a^3 = 0.37$, $b/a = 0.98$, and $a = 410 \pm 20$ $\mu$m); morphology B ($V/a^3 = 0.30$, $b/a = 0.43$, and $a = 392 \pm 20$ $\mu$m); and morphology A ($V/a^3 = 0.08$, $b/a = 0.61$, and $a = 512 \pm 16$ $\mu$m). Results of the numerical simulations considering the lyophilic contact angles $\theta_2 = 15°$ (blue) and $\theta_2 = 33°$ (red). The black curve in the radius plot shows the contour of the elliptical patch.

## ■ RESULTS

In this section, we present the results of our experiments and the Surface Evolver calculations for each of the observed morphologies. In Figure 5, each row presents one of the morphologies shown previously in Figure 4. For each morphology, we show the normalized footprint radii $R/a$ next to the local contact angles $\theta$, both along the azimuthal coordinate $\phi$. In each of the plots, we overlay experimental and computational results. The red and blue markers represent the calculations done considering the lyophilic contact angles $\theta_2 = 33°$ and $\theta_2 = 15°$, respectively. In the plots of the normalized footprint radii, we plot the position of the patch contour (black curve).

To follow the chronological sequence of our evaporating experiments, we present the results starting from morphology

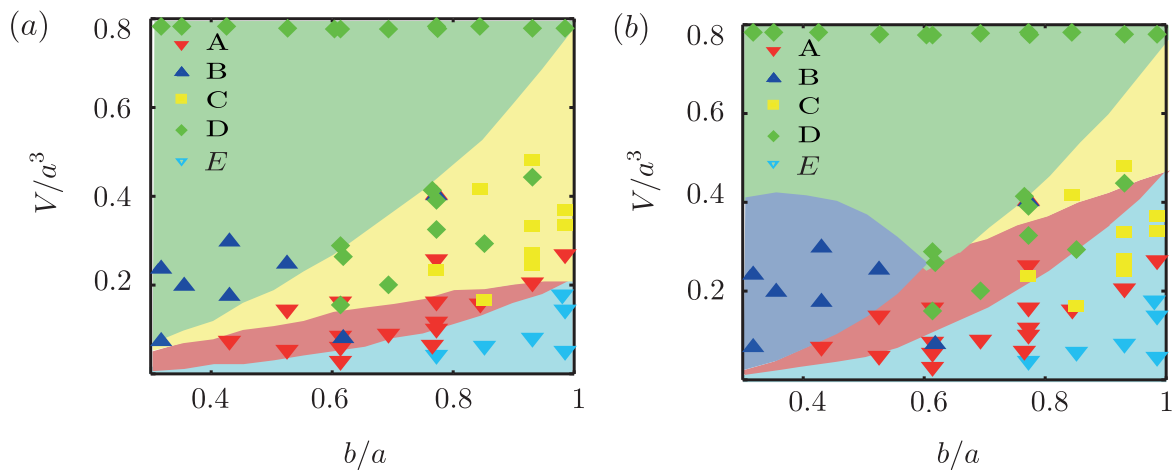

**Figure 6.** Morphology diagrams in aspect ratio $b/a$ vs relative volume $V/a^3$ phase space showing the morphologies A, B, C, D, and E. (a) Experimental results displayed together with the computational results considering $\theta_1 = 49°$ and $\theta_2 = \theta_{a2} = 15°$ (b) Experimental results displayed with the computational results considering $\theta_1 = 49°$ and $\theta_2 = \theta_{a2} = 33°$. The main features for A–D are indicated in Figure 4, while E shows the case in which the droplet is small enough to adopt the trivial spherical cap shape inside the patch. The color shadowed regions represent the morphologies obtained with our calculations. Green, yellow, dark blue, red, and light blue regions represent, respectively, the regions where morphologies D, C, B, A, and E were found in our calculations, and the colored markers show the experimental points specified in the legend.

D (largest droplet volume) and finishing with morphology A (smallest droplet volume). Finally, in Figure 6 we present the morphology diagrams predicted by the Surface Evolver calculations (colored areas) and compare those with the results of all our experiments (colored markers).

**Morphology Type D.** In Figure 5a,b, we can see an example of an experiment in which morphology D was found. The normalized footprint radius $R/a$ and the contact angle are plotted along the azimuthal coordinate. From the figure, we observe reasonable agreement between Surface Evolver calculations and experiments. However, the existence of inhomogeneities introduces pinning and hence, a delay in the movement of the contact line. This delay translates into an experimental radius larger than that predicted by the Surface Evolver calculations. Despite this mismatch of the contact line, which is a consequence of individual pinning events of the contact line, we obtain reasonable agreement with the contact angle calculations.

**Morphology Type C.** Morphology C appeared for ellipses of higher aspect ratio $b/a$ as compared to those of morphology B. Additionally, when being evaporated, the drop reaches morphology C always through morphology D, implying that any pinning event of the contact line outside the ellipse prevents morphology C. Indeed, the transition from morphology D to C was always found in a later stage of evaporation than that predicted by theory, i.e., for lower volumes (see Figure 6b).

In Figure 5c,d, we show a droplet of volume $V/a^3 = 0.37$ placed on a high aspect ratio ellipse ($b/a = 0.98$). The results of the calculations predict morphology C and contact angles between the receding and the advancing ones. The contact angle and the contact line show always good agreement with the Surface Evolver calculations. This is expected as it is the closer case to the trivial spherical cap shape.

**Morphology Type B.** The experimental data shown in Figure 5e,f are particularly interesting as this case presents a strong asymmetry in the radius caused by a sharp pinning point which can be identified at $\phi \approx 120°$ (indicated by an arrow in Figure 5e). Unlike the radius, the contact angle shows a symmetric behavior. We found that pinning leads to an asymmetric behavior in this morphology for all our experiments. However, besides the asymmetry forced by pinning, the experiments agree with the Surface Evolver calculations. The good match for the angle can be explained considering that the contact angle at every point of the contact line—far enough from the ellipse contour— is dictated by the chemistry of the surrounding substrate.

**Morphology Type A.** The experimental results showed morphology type A as predicted by the calculations for $\theta_2 = 15°$, with a part of the contact line pinned at the boundaries of the ellipse and the rest of the contact line inside the ellipse. Note that in the calculations for the higher receding contact angle ($\theta_2 = 33°$), the results predict a spherical cap shape with radius $R < b$. However, our results for the contact angle were not in good agreement with either of the Surface Evolver calculations but rather with an intermediate state between them, subjected to the irregularities of the edge (see Figure 5g,h). This can be due to imperfections of the coating in the edges of the ellipse. The high portion of the contact line that remains pinned at the boundary between the lyophilic and the lyophobic parts appeared to be very sensitive to the quality of the patch rim.

**Morphology Diagrams.** Figure 6 presents two morphology diagrams showing all the morphological regions and transitions predicted by the Surface Evolver calculations (as color shaded areas), together with the experimental results (colored markers). In this figure, morphology E is added to illustrate the transition to the case in which the ellipse does not have an effect, as in that case, the droplet is smaller than the ellipse minor axis. Using the receding contact angle for the calculations (see Figure 6a), is, in principle, the most logical method for calculating the shape of evaporating droplets. However, the experimental results show a behavior that falls between the results calculated for both limits of contact angle hysteresis.

In fact, for the first transitions (from morphology D to B and to C), the calculations done considering the hysteresis limits $\theta_{r1} = 49 \pm 4°$ and $\theta_{a2} = 33 \pm 4°$ show better qualitative agreement with our experiments than those done considering both receding angles. For these morphologies, in which part of the contact line lays on the lyophobic area, the overall angle of the drop is kept higher by the influence of the high contact angle of the lyophobic area $\theta_1 = \theta_{r1} = 49 \pm 4°$, bringing it to the maximum angle possible for the lyophilic patch $\theta_2 = \theta_{a2} = 33 \pm 4°$. For the same reason, the transitions from morphologies B and C to morphology A, in which the contact line has to travel along the lyophilic patch, show better agreement with the calculations performed for $\theta_2 = \theta_{a2} = 15 \pm 4°$. The experiments also show that the transitions occur for smaller droplet volumes as compared to those predicted. This delay can be observed if, for a fixed aspect ratio $b/a$, we follow down the vertical line in the decreasing volume direction. This effect is forced by pinning, which acts to favor larger radii morphologies.

Moreover, during the calculations we found that, for certain lyophilicity differences, we can exclude morphologies from the diagram. In our case, the calculations that were computed considering $\theta_2 = \theta_{r2} = 15°$ (see Figure 6a) predict the absence of morphology B.

## CONCLUSIONS

We have performed experiments to validate the Surface Evolver calculations, and these showed good agreement. We observe how the effect of substrate heterogeneities, including pinning and contact angle hysteresis, can affect the accuracy of our surface minimization calculations. These heterogeneities seem mostly to affect the symmetry and the transitions in the morphology diagrams. With this study, we show the robustness of the contact angle predictions which contrast with the sensitivity that radii predictions have to pinning. The reason is that the radius depends on the mobility of the contact line and therefore on pinning, even when a different morphology would be energetically more efficient, while the contact angle is forced by the chemistry of the substrate in the vicinity of the contact line, making it more robust.

According to our results, we conclude that the knowledge of the hysteresis limits can be used to improve the predictions of Surface Evolver calculations. In general, the morphologies that were found experimentally show good repeatability. However, the morphologies adopted by the droplets are always subjected to the effects of pinning, which influence the droplet's symmetry and delays its transition to the next morphology. This effect shifts the experimental transitions to smaller volumes than those predicted by our calculations, as shown in the morphology diagram. We expect this effect to be the opposite for growing droplets, but that remains an open question, and it is beyond the scope of the present study, as it would require a different experimental setup. Finally, the exploration of lyophilicity differences between the patches and the surroundings has shown the feasibility of excluding morphologies from the phase diagram, which is an interesting result with bearing on the design of micropatterned structures for various applications.

## AUTHOR INFORMATION

**Corresponding Authors**

*E-mail: d.lohse@utwente.nl.

*E-mail: xuehua.zhang@ualberta.ca.

**ORCID**

José M. Encarnación Escobar: 0000-0002-2527-7503

Ivan Dević: 0000-0003-0977-9973

Xuehua Zhang: 0000-0001-6093-5324

Detlef Lohse: 0000-0003-4138-2255

**Author Contributions**

#Both authors contributed equally to this work

**Notes**

The authors declare no competing financial interest.

## ACKNOWLEDGMENTS

This work was supported by The Netherlands Center for Multiscale Catalytic Energy Conversion (MCEC), an NWO Gravitation program funded by the Ministry of Education, Culture and Science of the government of The Netherlands. D.L. also acknowledges an ERC Advanced grant. We acknowledge the support of the Natural Sciences and Engineering Research Council of Canada (NSERC).